# First-principles prediction of Structural Stability and Thermoelectric Properties of SrGaSnH

Enamul Haque*[1] and Mizanur Rahaman[2]

[1]EH Solid State Physics Laboratory, Longaer, Gaffargaon, Mymensingh-2233, Bangladesh

[2]Department of Physics, Mawlana Bhashani Science and Technology University, Santosh, Tangail-1902, Bangladesh.

*Email: enamul.phy15@yahoo.com

**Abstract**

Thermoelectric materials based on earth-abundant and non-toxic elements are very useful in cost-effective and eco-friendly waste heat management systems. The constituents of SrGaSnH are earth-abundant and non-toxic, thus we have chosen SrSnGaH to study its structural stability and thermoelectric properties by using DFT, DFPT, and semi-classical Boltzmann transport theory. Our elastic and phonons calculations show that the compound has good structural stability. The electronic structure calculation discloses that it is an indirect bandgap (0.63 eV by mBJ+SOC) semiconductor. Light band hole effective mass leads to higher electrical conductivity along x-axis than that of along z-axis. On the other side, the weak phonon scattering leads to high lattice thermal conductivity ~10.5 W m$^{-1}$K$^{-1}$ at 300 K. Although the power factor (PF) is very high along the x-axis (above 10 mW m$^{-1}$K$^{-2}$ at 300 K), such large $\kappa_l$ dramatically reduces ZT. The maximum values of in-plane and cross-plane ZT are ~1 (n-type), 0.8 (p-type) and 0.6 (n-type), (0.2 p-type) at 700 K, respectively. The present study has revealed that this compound has strong potential in eco-friendly TE applications.

## 1. Introduction

In the past decades, hydrogen-based materials have been extensively studied due to their practical applications in different types of systems [1–4]. For example, some hydrides show an extraordinary superconducting property under pressure, such as ThH$_{10}$ [5,6]. But the suitability of hydrogen-based compounds in thermoelectric device applications has not been established yet.

Thermoelectric (TE) materials can generate electricity from heat and thus, they are clean energy resources [7–9]. The efficiency of TE materials can be estimated from a dimensionless quantity, called ZT and defined [10,11] by $ZT = \frac{S^2 \sigma}{\kappa} T$, where S, σ, T, and κ are the Seebeck coefficient, electrical, absolute temperature, and thermal conductivity, respectively. Generally, a large bandgap and effective mass cause a high Seebeck coefficient but low electron conductions and vice versa. On the other hand, some materials have intrinsic high thermal conductivity (composed of electrons and phonons part). Therefore, it is difficult to optimize these parameters simultaneously. Although there have been proposed different techniques (such as alloying and nanostructuring [12–14]) to optimize them, none can successfully optimize them to the optimal point. However, these methods can improve thermoelectric performance significantly. Thus, it is still challenging to obtain a high ZT value (>3) for commercial TE device applications. Generally, strong phonon scattering is induced either in the compounds with a very complex structure or containing heavy elements/halogens. The lattice thermal conductivity in these systems is usually low while the power factor remains high. Such types of materials are well desired and researchers around the globe have been searching these materials since the discovery of the TE effect [15].

Recent theoretical and experimental studies on the thermoelectric performance of Zintl phases have revealed that these phases possess tremendous potential in TE applications [16–18]. In Zintl phases, the alkaline earth metal donates electrons to covalent-bonded layers of the anionic metalloids (or post-transitions metals). The anionic layers cause significant band overlapping and light effective mass, which leads to high carrier transport. On the other hand, cations are suitable for alloying without affecting anionic layers significantly. Moreover, Zintl phases usually have an optimum bandgap [19]. These features are responsible for the high TE performance of the Zintl phases. The mixing of elements from groups I and II (metals and metalloids) can form a family of compounds with a well-known structure [20]. A family of compounds EmTmMt (Em= Alkaline earth metals; Tm= post-transition metals (Al, Ga, Sn); and Mt = metalloids (Si, Ge)) are metallic conductors [20]. In these compounds, highly electronegative elements interact (termed as polyanion) with other atoms and the mixing of hydrogen induces closed-shell electronic configuration, converting them into polar semiconducting materials with chemical formula EmTmMtH [21–25]. The bandgap of these compounds was theoretically predicted to be between

0.3-0.8 eV, while the bandgap of SrAlSiH was measured experimentally to be 0.63 eV [25]. The reported electronic structures suggest that conduction band minima (CBM) and valence band maxima (VBM) of these compounds are highly dispersive, while CBM is a non-degenerate band but VBM is degenerate [23]. The dispersive and degenerate characters of band favors high Seebeck coefficient and electrical conductivity, leading to high power factor. In general, a small bandgap semiconductor with these features shows good thermoelectric performance, but thermoelectric properties of this compound have not been investigated yet.

Here, we report the details of the electron-phonon scattering mechanism and thermoelectric performance of SrGaSnH. Our first-principle calculations reveal that phonons are weakly scattered in SrGaSnH, leading to relatively high lattice thermal conductivity. On the other side, light effective mass leads to high electrical conductivity and hence high power factor. Our investigation predicts that these compounds have a strong potential in thermoelectric device applications.

## 2. Computational Details

In this current work, we used two density functional theory (DFT) codes. First, we fully relaxed the structural parameters by minimizing interatomic forces below 0.1 eV/A in WIEN2k, a full-potential linearized augmented plane wave method (FP-LAPW) based DFT code [26]. We set muffin tin (Rmt) sphere radii to 1.95, 1.93, 2.1, and 1.34 Bohr for Sr, Ga, Sn, and H, respectively, and kinetic energy cutoff RKmax=5.0 (which determines the size of the plane wave expansions) due to small sphere radii of H, after extensive trials. In this calculation, we used generalized gradient approximation (GGA) with PBEsol setting [27,28] and $10 \times 10 \times 8$ k-point. By using the relaxed structure, we performed elastic properties [29] (using the same setting) and electronic structure calculations. As GGA usually underestimates bandgap severely, we used Tran-Blaha modified Becke Johnson potential (mBJ) [30,31] including spin-orbit coupling and $30 \times 30 \times 30$ k-point. The obtained energy Eigenvalues and electronic parameters from the above self-consistent calculations (SCF) were then fed into a modified BoltzTraP code [32]. To calculate the carrier relaxation time, we performed phonon calculations by using density functional perturbation theory (DFPT) implemented in Quantum Espresso (QE) [33], a plane wave-based DFT codes. To proceed this, we again fully relaxed the structure in QE by using 6 6

4 k-point, Vanderbilt ultrasoft pseudopotential [34] with a cutoff energy 46 Ry and Marzari-Vanderbilt smearing (width 0.03 Ry) [35], and strict convergence criteria (energy threshold=$10^{-14}$ Ry and force=$10^{-5}$ Ry/Au). By using the fully relaxed structure, we performed electron-phonon calculations with $10 \times 10 \times 8$ k-point and 444 q-point. The dynamical matrix was then fed into EPA code [36] to obtain an average electron-phonon dynamical matrix, by using an energy bins-based method. In this case, we used 6 bins with 0.5 eV energy intervals. This average matrix was then fed into modified BoltzTraP code along with other required parameters to calculate carrier relaxation time.

For lattice thermal conductivity calculations, we used the finite displacement method as implemented in ShengBTE [37]. For this, we created 222 supercells in Phononpy code [38] to calculate second-order force constants (IFCs). For third-order IFCs, we created the supercells with the same size by using thirdorder_espresso.py script supplied with ShengBTE. We computed interatomic forces in QE by using the same setting described above except the k-point in this case, $2 \times 2 \times 2$ k-point). The lattice thermal conductivity was then calculated (by using the obtained second order and third order IFCs) by solving the phonon Boltzmann transport equation (BTE) within the relaxation time approximation (RTA) and full-iterative solution method (with $16 \times 16 \times 16$ q-point). This method has been found very successful in predicting the lattice thermal conductivity of many materials [39–41].

## 3. Results and Discussion

SrGaSnH, a hydrogen-based Zintl phase, forms a Hexagonal unit cell (four atoms and formula unit Z=1) [21]. Its space group symmetry is P3m1 (#156). In this symmetry, each element Sr, Ga, Sn and H occupies the 1a (0, 0, 0), 1c (2/3, 1/3, 0.6), 1b (1/3, 2/3, 0.45), and 1c (2/3, 1/3, 0.94) the Wyckoff positions, respectively. The Ga atom is bonded with the H atom, as shown in Fig. 1.

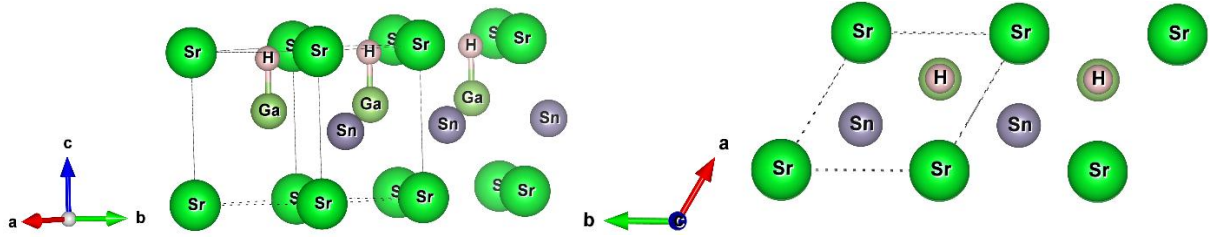

Fig. 1. The unit cell of SrGaSnH and its side view. The structure has been extended along [010] direction.

The distance between H and Ga is 1.77 Å, which is slightly longer than that between H and Al in CaAlSiH (1.75 Å) [42]. Note that Sr atom is also bonded with H atom (not shown in Fig. 1) with a distance of 2.24 Å. The concept of bonding in this family of Zintl phases can be understood from the valence electrons configurations. The state can be formulated in two ways either as $(Sr)^{2+}[GaSnH]^{2-}$ or $[SrGaSn]^{1+}(H^{1-})$ [23,25,43]. In the first configuration, Ga-Sn covalently bonded layer is the electron acceptor and Sr is the electron donators, while Sn has a lone pair electrons. The second configuration suggests that H-1s orbital has negligible contributions near the Fermi level and the main contributions are about 5 eV below the Fermi level [23]. In that case, a weak hybridization among H-1s, Ga-4p, and Sn-5p orbitals exists near the Fermi level. These configurations consider the oxidation state of Sr, Ga, Sn, and H to be $2^+$, $3^+$, $4^-$, and $1^-$, respectively. Note that the first configuration describes the SrGaSnH as the Zintl phase and a more powerful concept.

### 3.1. Structural stability

As SrGaSnH is not synthesized yet, it is necessary to assess its structural stability before proceeding to our main discussion. The formation enthalpy, taking the difference between total energy and some of the energy of each constituent atoms, can predict the energetic formability when we mix the constituents at the required conditions.

Table I: Calculated lattice parameters, elastic moduli (in GPa), Pugh's (B/G [44]) and Poisson ratio (v) [45], longitudinal ($v_l$), transverse ($v_t$), and average sound velocity, Debye temperature ($\theta_D$) [46], and melting temperature ($T_m$) [47].

| Parameters | $a(Å)$ | $c(Å)$ | $c_{11}$ | $c_{12}$ | $c_{13}$ | $c_{33}$ | $c_{44}$ | $B$ | $G$ |
|---|---|---|---|---|---|---|---|---|---|
| Values | 4.19 | 5.03 | 116 | 30 | 20 | 52 | 32 | 44 | 33 |
| Parameters | $Y$ | $B/G$ | $v(km/s)$ | $v_l(km/s)$ | $v_t(km/s)$ | $v_a(km/s)$ | $\theta_D(K)$ | $T_m(\pm 500K)$ | |
| Values | 85 | 1.3 | 0.2 | 4.05 | 2.49 | 2.74 | 294 | 1014 | |

Our computed value of this parameter is -0.49 eV/atom, suggesting that the compound is energetically favorable. Our computed lattice parameters, elastic constants, and other related parameters are listed in Table I. The calculated ground state lattice parameters fairly agree with other reported theoretical values by using PBE functional [21]. From the table, it is clear that the elastic constants of SrGaSnH satisfy the stability criteria, as described in Ref. [48]. Therefore, the compound is mechanically stable. The Debye temperature is much higher than that of good thermoelectric materials but relatively smaller than that of a few half-Heusler compounds [49].

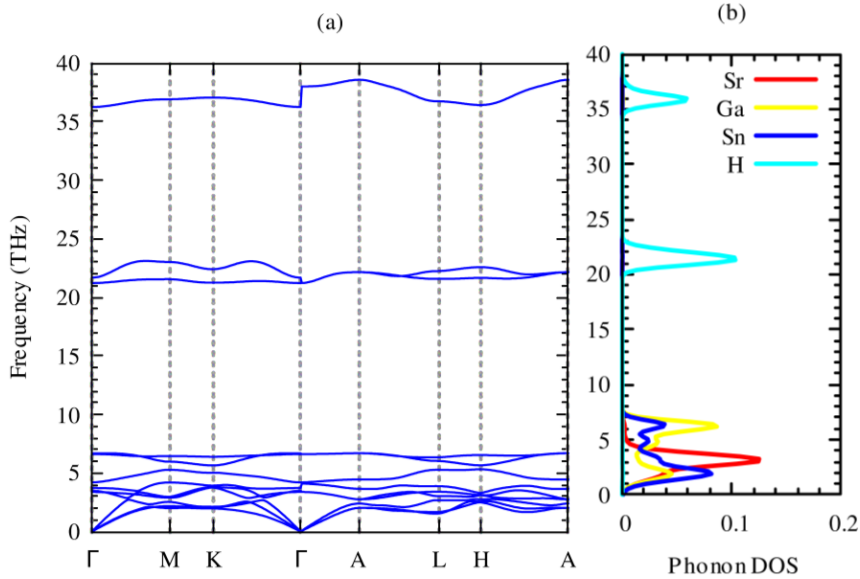

Fig. 2. Phonon dispersion relations and partial phonon density of states of SrGaSnH along with high symmetry points.

Unfortunately, the Pugh's ratio suggests that this compound is brittle. However, through doping with a suitable element, the compound may be converted into ductile type. The estimated melting temperature from elastic constants is comparatively smaller than that of Hal-Heusler [50]. Therefore, this compound cannot be used for high-temperature thermoelectric device applications. The dynamical stability is the most important criterion for a compound to be synthesized practically [51].

This assessment of these criteria for this compound can be predicted from the phonon dispersion relations. The phonon dispersion relation and partial phonon density of states of SrGaSnH are shown in Fig. 2. The positive phonon energy over the whole Brillouin zone indicates the dynamical stability of the compound. In our phonon calculations, we consider non-analytical correction by calculating macroscopic dielectric constant and Born effective charges [52], as listed in Table II. From Fig. 2, we see that longitudinal optical phonon and transverse optical phonon split (LO-TO splitting) at Γ-point. Both dielectric constant and effective charges show highly anisotropic behavior, suggesting a similar trend of carrier transport. From partial phonon density of states, we see that Sr, Ga, and Sn induce three acoustic phonons, where Sr has dominated contributions.

Table II: Computed in-plane and cross-plane macroscopic dielectric constant ($\varepsilon$) and Born effective charge ($Z^*$).

| $\varepsilon(x)$ | $\varepsilon(z)$ | | $Z^*(x)$ | $Z^*(z)$ |
|---|---|---|---|---|
| 38 | 14 | | | |
| | | Sr | 1.98 | 2.25 |
| | | Ga | 0.99 | 0.74 |
| | | Sn | -2.57 | -1.12 |
| | | H | -1.44 | -1.84 |

There is no gap between acoustic and optical phonons, which is non-conducive for heat, suggesting relatively moderate phonon scattering in this compound. The first six optical phonons of frequency ~120-200 cm$^{-1}$ arise from Ga and Sn, with a very small contribution of Sr. Three high energy optical phonons come from H, suggesting that H has negligible contribution in the lattice thermal conductivity.

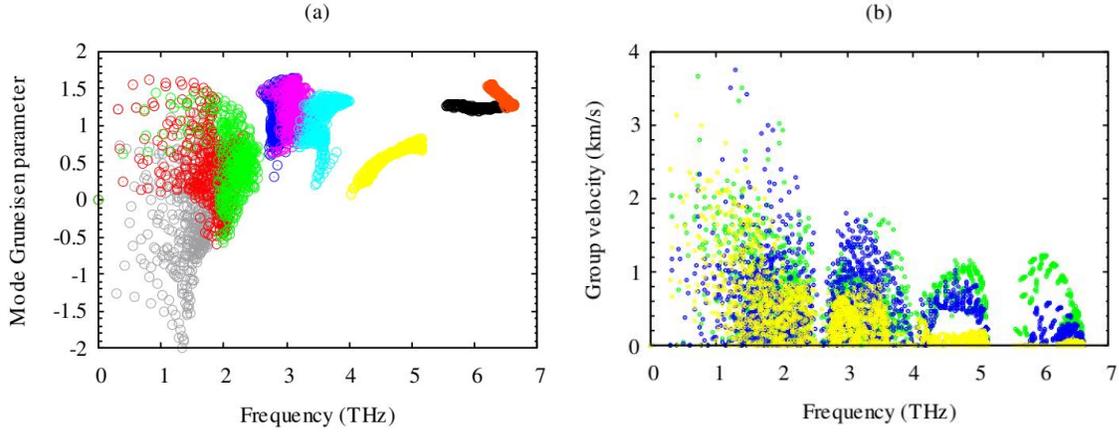

Fig 3. Computed phonon mode of Gruneisen parameter (a) and group velocity (b) from second and third-order IFCs of SrGaSnH. The high-frequency phonons of H are omitted here for clarity.

Lattice thermal conductivity is directly related to the phonon group velocity and inversely to the Gruneisen parameters. Our computed group velocity and mode Gruneisen parameters from third-order IFCs are illustrated in Fig. 3. The high group velocity of SrGaSnH suggesting relatively larger value of lattice thermal conductivity. On the other hand, a small (relatively) value of the Gruneisen parameter indicates moderate anharmonicity and thus, moderate phonon scattering, vice versa, in SrGaSnH. Now let us see the effect of such a weak phonon scattering on the lattice thermal conductivity. Fig. 4. (a) shows the calculated $\kappa_l$ at different temperatures. The $\kappa_l$ exhibits strong anisotropic behavior, as structural arrangement, suggests that phonons cannot propagate along z-direction easily as compared to the x-direction.

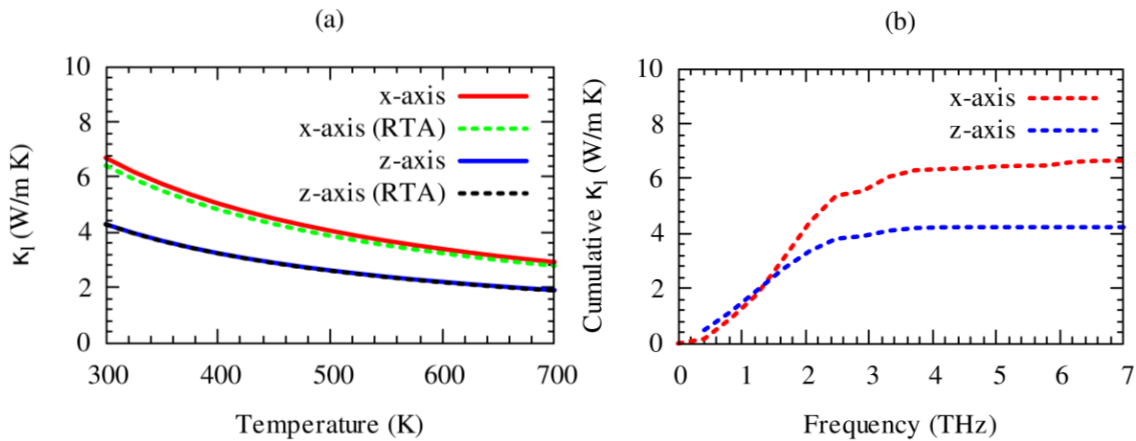

Fig. 4. (a) Temperature-dependent anisotropic lattice thermal conductivity by using two different methods (relaxation time approximation (RTA) and full iterative solutions of BTE (solid lines),

and (b) Cumulative lattice thermal conductivity by using full iterative solutions of BTE. High-frequency phonons of H are ignored here because they have a negligible contribution to the heat conduction.

The room temperature kl of SrGaSnH is comparable to that of some Half-Heusler compounds [49,53,54]. Interestingly, RTA [55] underestimates $\kappa_l$ slightly along x-direction only as compared to the full iterative solutions of Boltzmann transport equation (BTE) [56,57]. About 90% of lattice thermal conductivity comes from acoustic phonons, which can be seen from the cumulative lattice thermal conductivity as shown in Fig. 4. (b). Sr-atom has the dominant contribution to the acoustic phonons, thus, most of the heat is conducted through the acoustic phonons of Sr. This suggests that Sr is the suitable site for doping to reduce lattice thermal conductivity without affecting electronic structure significantly. In the next section, we will discuss the features of the electronic structure.

From the calculated Gruneisen parameter we see that the acoustic phonons are below zero and the negative value of the Gruneisen parameter indicated that the structural instability. The thermal conductivity is inversely proportional to the Gruneisen parameter. The low value of the Gruneisen parameter indicates high thermal conductivity and good in thermoelectric applications.

### 3.2. Electronic structure

Electronic structures signify the carrier transport of materials. Our computed band structure and projected density of states by using mBJ potential including the spin-orbit coupling effect are shown in Fig. 5. (a) and (b), respectively. The closed-shell electronic configuration of (18 electrons) SrGaSnH indicates that it should be a semiconductor. The conduction band minima (CBM) and valence band maxima (VBM) lie two different high symmetry points (at K and A), suggesting the indirect nature of bandgap.

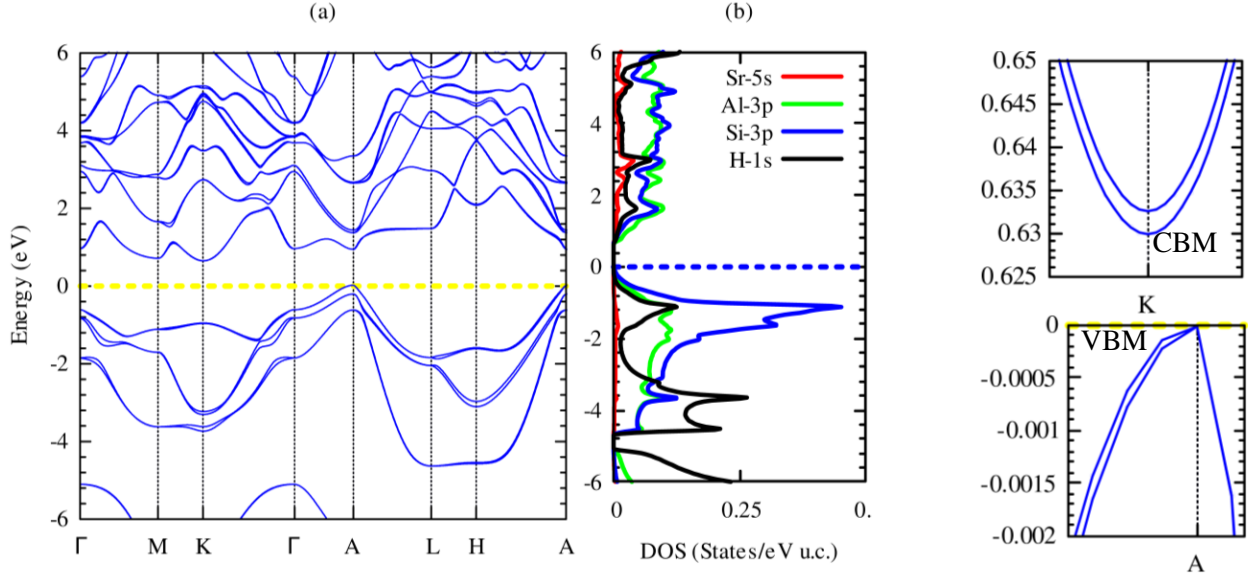

Fig. 5. Electronic band structure (a) and projected density of states (b) of SrGaSnH. The dashed lines at zero energy represent the Fermi level. The right panel shows the conduction band minima (CBM) and valence band maxima (VBM) of SrGaSnH.

Unlike SrAlSiH (CBM lies at M-point), the CBM of SrGaSnH is shifted to K-point. Furthermore, the PBE-functional without SOC pushes the CBM at A-point, converting it into a direct bandgap semiconductor, which is consistent with one reported in the materials project (ID: mp-978852). Our computed density of states is also consistent with the one reported in Ref. 16. Note that the computed bandgap value (0.63 eV) by using mBJ+SOC is lightly higher than that obtained by PBE functional (0.59 eV) without SOC [21]. This is expected because SOC usually reduces the bandgap by pushing the Fermi level upward, while PBE functional underestimates the bandgap about 40-50% by the experimental value. From the projected density of states, we see that Ga-4p and Sn-5p have the dominant contribution to the electronic band structure near Fermi level, while ha negligible contribution. On the other hand, H-1s has dominated contribution below -3 eV and major below -5 eV, as predicted by the second model of the Zintl phase. Interestingly, although CBM is a parabolic band arising mainly from Ga-4p, VBM is a non-parabolic band induced from Sn-5, as shown in the right panel of Fig. 5. On the other side, CBM is a non-degenerate band but VBM is a two-fold degenerate band with a single band extrema at A-point. However, both CBM and VBM are highly dispersive bands. By fitting third-order polynomial, we have calculated the effective mass of CBM and VBM. The effective mass of electrons is 0.11 m0, while light hole (light band (band #8)) and heavy holes (heavy band (band #9) masses are 0.08 and 0.4 m0. Therefore, the light hole may lead to high electrical conductivity, while the heavy effective mass of electrons (compared to light hole) may induce a high Seebeck coefficient. Besides, the band edge of CBM is slightly flatter than that of VBM, which will also be favorable for the high Seebeck coefficient. Let us visualize this prediction in the next section.

### 3.3. Transport Coefficients

Boltzmann transport equation implemented in BoltzTraP uses constant relaxation time approximation (cRTA) to evaluate transport coefficients. Thus, we need to evaluate carrier relaxation time explicitly. This can be done by using average electron-phonon approximation as implemented in EPA code [36]. In this approximation, the carrier relaxation time is calculated by using the following expression [36]

$$\tau^{-1}(\epsilon,\mu,T) = \frac{2\pi\Omega}{g_s\hbar}\sum_{v}\{g_v^2(\epsilon,\epsilon+\bar{\omega}_v)[n(\bar{\omega}_v,T)+f(\epsilon+\bar{\omega}_v,\mu,T)]\,\rho(\epsilon+\bar{\omega}_v) \\ + g_v^2(\epsilon,\epsilon-\bar{\omega}_v)[n(\bar{\omega}_v,T)+1-f(\epsilon-\bar{\omega}_v,\mu,T)]\rho(\epsilon-\bar{\omega}_v)\}\ldots\ldots(1)$$

where these symbols have usual meaning ($\Omega$ is the unit cell (primitive) volume, $\hbar$ is the reduced Planck's constant, $v$ the index of phonon modes, $\bar{\omega}_v$ the averaged frequency of phonon modes, $g_v^2$ the averaged of the electron-phonon matrix, $n(\bar{\omega}_v,T)$ the Bose-Einstein distribution function, $f(\epsilon+\bar{\omega}_v,\mu,T)$ the Fermi-Dirac distribution function, $g_s = 2$ the degeneracy of spin, $\epsilon$ the energy of electrons, and $\rho$ the DOS per unit volume and energy).

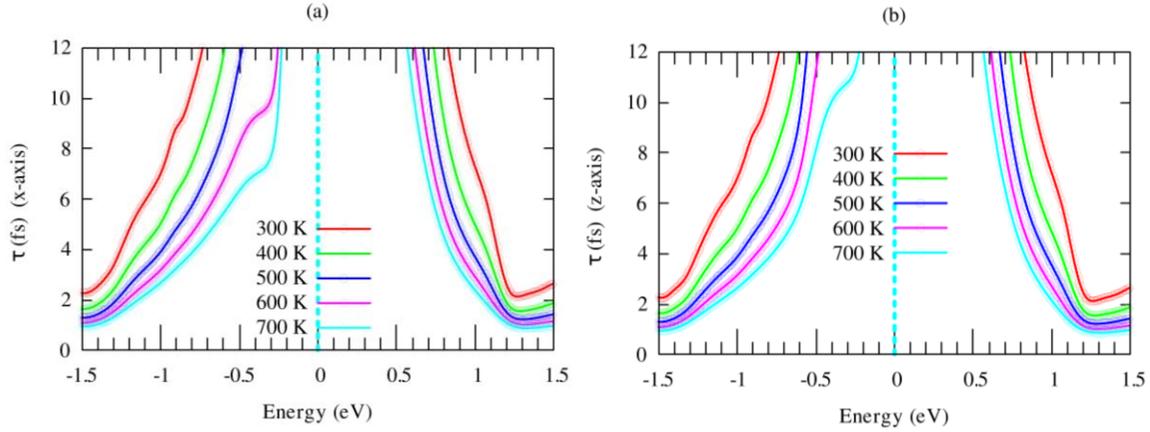

Fig. 6. Energy-dependent carrier relaxation time of SrGaSnH at different temperatures along (a) x- and (b) z-direction. The dashed lines represent the Fermi level.

We see that τ decreases rapidly around the band-edges. This behavior may be explained by using the following expression [36]:

$$\tau^{-1} \sim g^2(\epsilon)\rho(\epsilon) \ldots \ldots \ldots (2).$$

Therefore, the carrier relaxation time (τ) is inversely proportional to the carrier density of states (*ρ*) per unit energy and unit volume, but the electron-phonon matrix elements (*g*) show a weak carrier energy dependency.

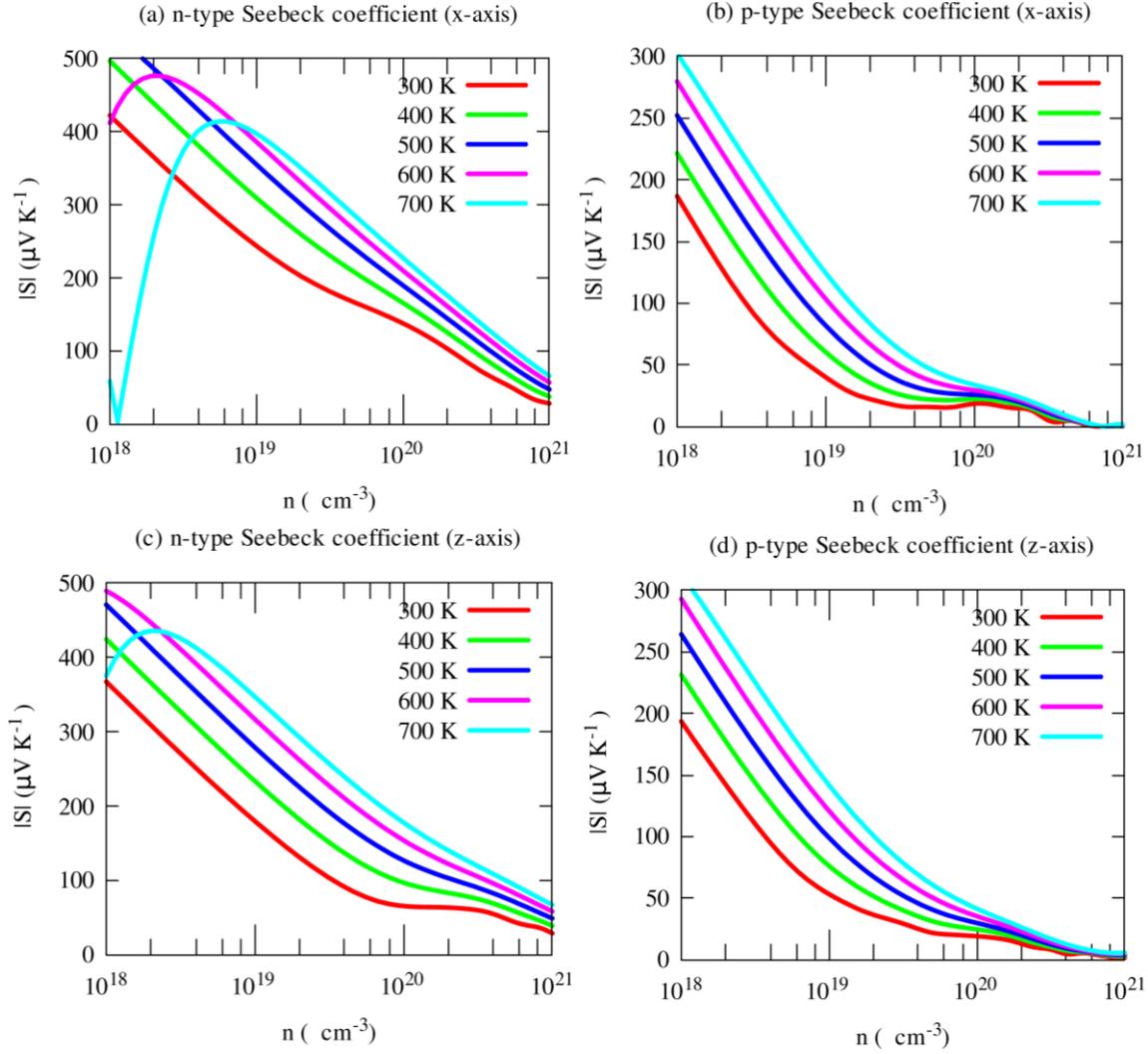

Fig. 7. Anisotropic Seebeck coefficient as a function of carrier concentration at different temperatures.

Seebeck coefficient of SrGaSnH exhibits a weak anisotropic behavior, as shown in Fig. 7. The n-type S is much higher than that of the p-type carrier, due to heavy effective mass and less dispersive bands compared to that of p-type carriers. It decreases monotonically with carrier density as usual, but not linear at all temperatures. The nonlinearity arises from the bipolar conduction effect. The n-type S is comparatively much higher than that of some half-Heusler compounds, such as TaCoSn [58] and HfCoSb [53,59].

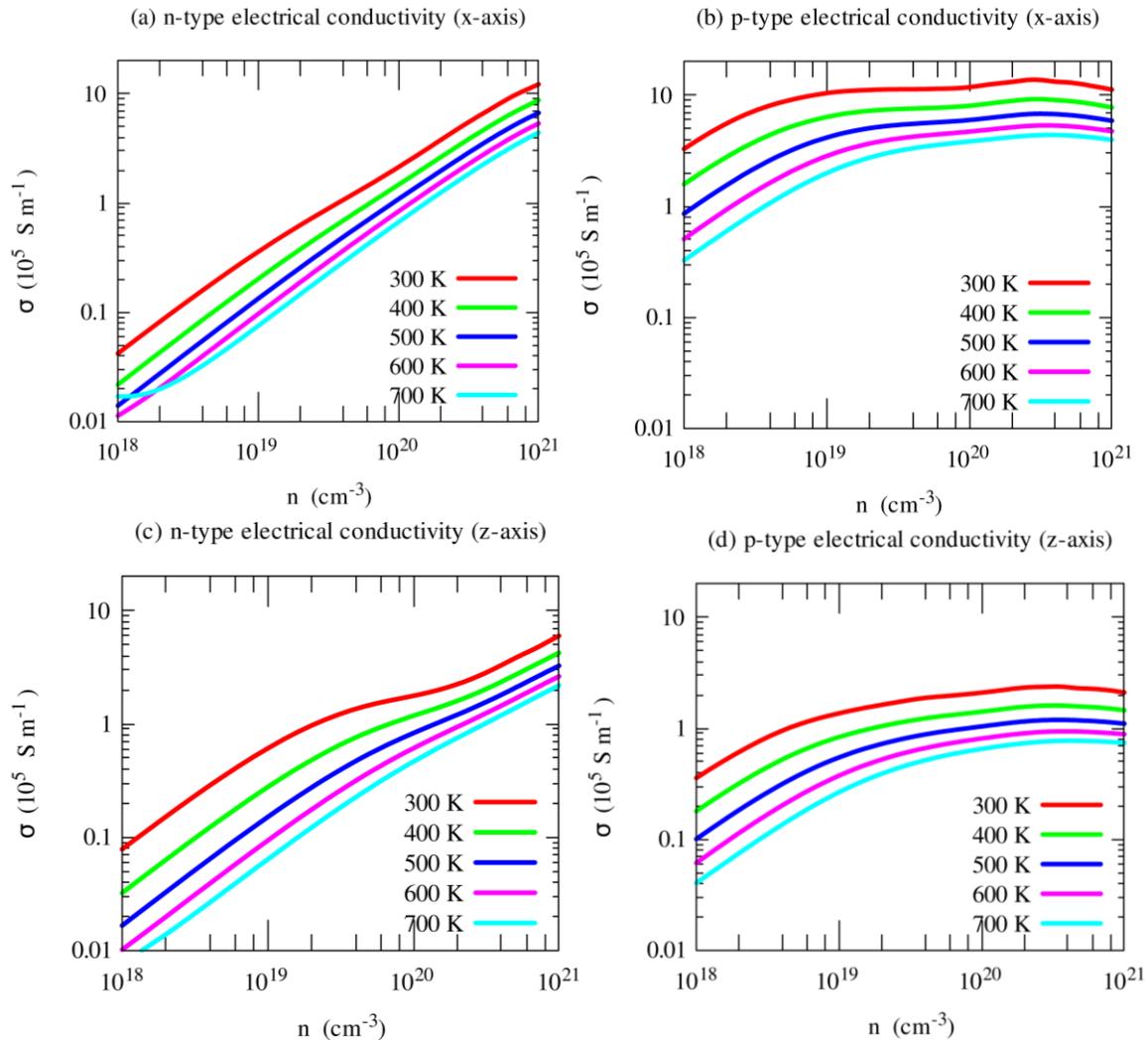

Fig. 8. Electrical conductivity along with crystallographic directions of SrGaSnH as a function of carrier density at different temperatures.

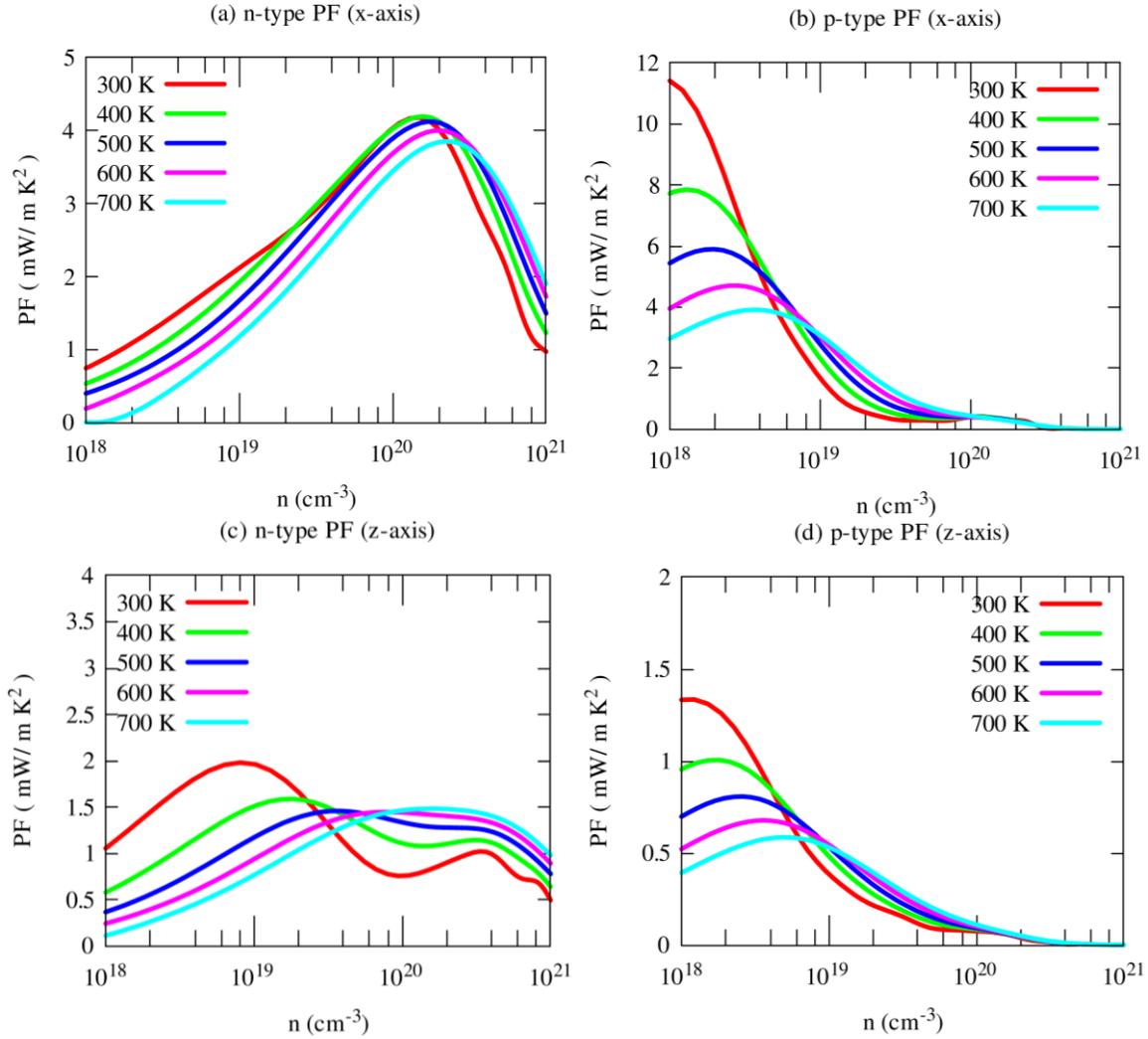

Fig. 9. Calculated in-plane (x) and cross-plane (x) power factor (PF) of SrGaSnH as a function of carrier concentration at the selected temperatures.

The electrical conductivity shows opposite trends. Light holes lead to much higher electrical conductivity than electrons, as shown in Fig. 8. Moreover, it shows strong anisotropic behavior. In-plane (x-axis) electrical conductivity of p-type carriers is much larger than that of along cross-plane. Unlike S, it increases with carrier density.

By the combination of high Seebeck coefficient and electrical conductivity, the computed power factor of SrGaSnH is large, reaching ~11 $mW/mK^2$ along x-direction at 300K for the p-type carrier. But cross-plane power factor is much smaller for both n- and p-type carriers. Such a high in-plane power factor of SrGaSnH suggests that this compound has strong potential in thermoelectric device applications.

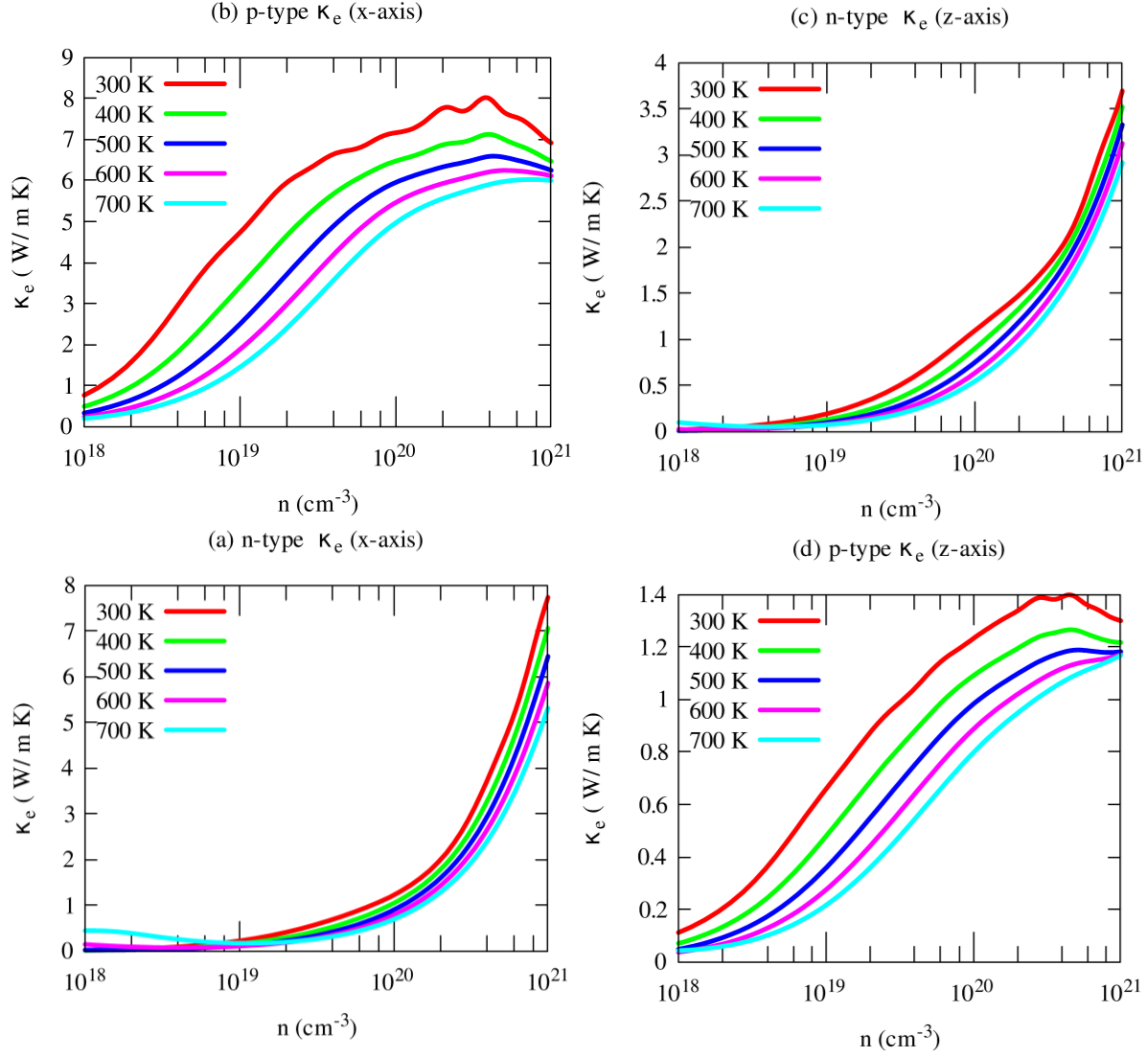

Fig. 10. The anisotropic electronic part of the thermal conductivity of SrGaSnH at five consecutive temperatures.

However, we still need to assess two other quantities for final prediction, namely, the electronic part of the thermal conductivity and thermoelectric figure of merit (ZT). Fig. 10. Illustrates the variations of electronic thermal conductivity with carrier density at different temperatures. The electronic contribution to the heat conduction shows similar trends of electrical conductivity. Like electrical conductivity, the in-plane electronic part of the thermal conductivity for the p-type carrier is also much higher along than that of other axes and carriers. In this case, both carriers and phonons almost equally contribute to the heat conduction. However, in other cases, phonons dominantly contribute to the heat conductions, especially for n-type carriers along both axes. This indicates that the reduction of lattice thermal conductivity may improve its thermoelectric performance effectively.

Table III: Evaluated room temperature transport properties of SrGaSnH (values in the bracket correspond to the cross-plane data)

| Carrier type | n ($10^{19}$ cm$^{-3}$) | S (μV/K) | τ(ps) | σ ($10^5$ S/m) | $\kappa_e$(W/m K) | PF(mW/m K$^2$) |
|---|---|---|---|---|---|---|
| n | 7.9 | 147 (68) | 0.022 (0.03) | 1.77 (1.66) | 1.06 (0.97) | 3.82 (0.76) |
| p | 0.15 | 150 (162) | 1 (0.9) | 4.5 (0.49) | 1.22 (0.17) | 10.1 (1.29) |

### 3.4. Thermoelectric performance

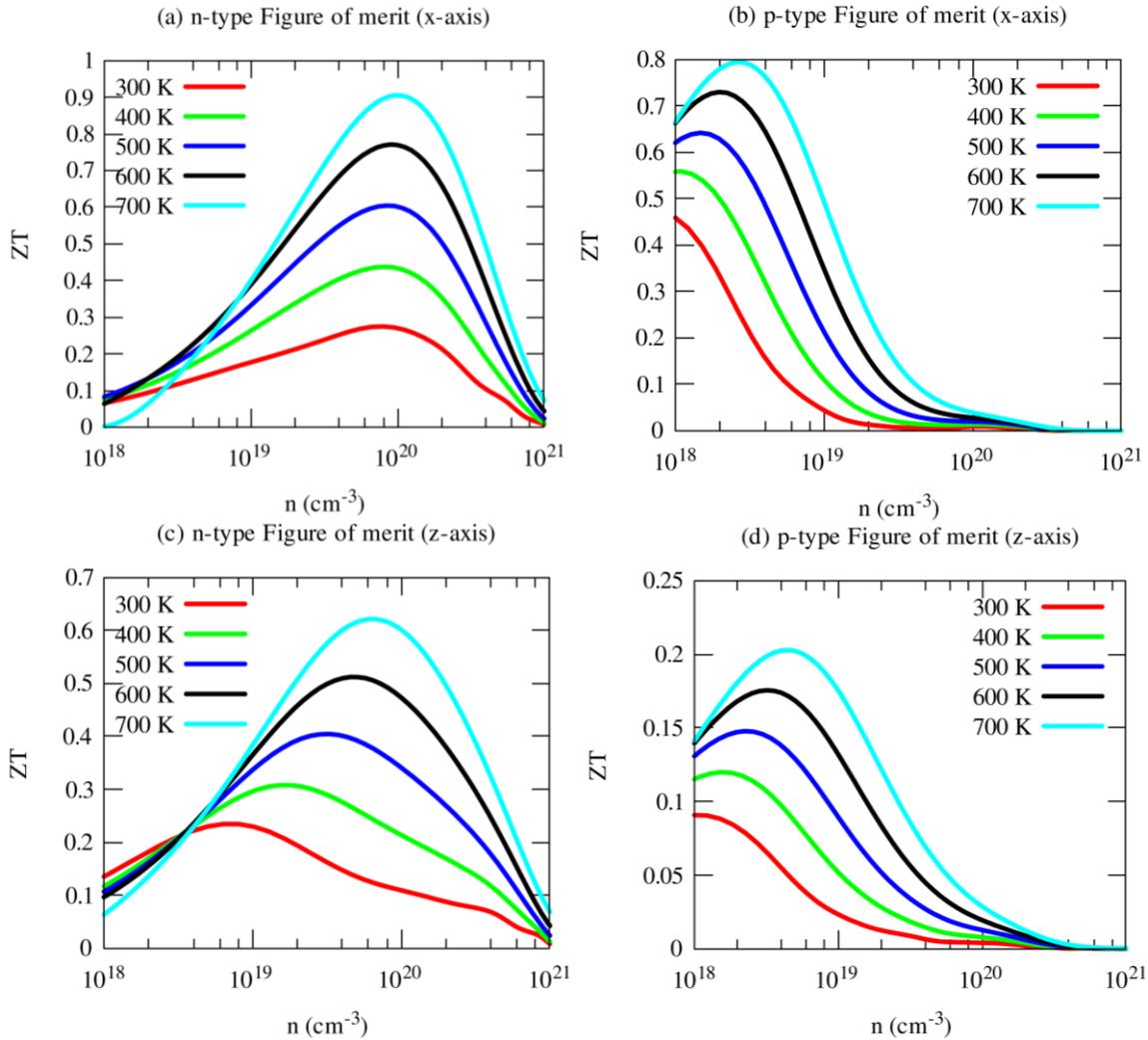

Fig. 11. Evaluated anisotropic thermoelectric figures of merit (ZT) at five consecutive temperatures.

Now we have obtained all quantities to calculate its thermoelectric figure of merit (ZT). Our computed ZT of pristine SrGaSnH is relatively high, reaching ~1.0 at 700K (and $n \sim 8 \times 10^{19}\ cm^{-3}$) for n-type carriers along the x-axis, as shown in Fig. 11. In-plane ZT of p-type carriers is smaller compared to that of n-type carriers, although the p-type PF is two times higher than n-type, due to high electronic thermal conductivity. The cross-plane ZT of n-type SrGaSnH reaches to 0.6 at 700K, while it remains much lower for p-type. Usually, the lattice thermal conductivity may be reduced dramatically through alloying or nano-structuring without affecting the electronic structure significantly. Therefore, the ZT would be then much higher and SrGaSnH would be practical materials for thermoelectric device applications for low-medium range temperatures.

## 4. Conclusions

In summary, we have performed a series of first-principles calculations to study structural stability and thermoelectric performance of SrGaSnH. From formation energy and elastic constants, we find that the compound under consideration is energetically and mechanically stable, suggesting that it is favorable to synthesize in the laboratory. Furthermore, lattice dynamic calculations indicate that the compound is dynamically stable and relatively weak phonons scattering, leading to high lattice thermal conductivity, 6.7 and 4.3 W/m K (in-plane and cross-plane, respectively) at 300 K. On the other hand, relatively light effective mass of holes and electrons leads high electrical conductivity along the x-direction, especially light hole leads to an extraordinary in-plane power factor ~11 mW/m K$^2$ at 300 K. However, the thermoelectric performance is dramatically reduced due to high lattice thermal conductivity and in-plane electronic part of the thermal conductivity (p-type). Thus, the TE performance may be improved substantially through alloying with a suitable element or nanostructuring, without affecting the electronic structure significantly. The calculated ZT of SrGaSnH is ~1.0 (n-type), 0.8 (p-type) and 0.6 (n-type), 0.2 (p-type) along x- and z-directions. Therefore, the compound has strong potential in TE device applications operated at low-medium range temperatures. We hope that experimentalists will synthesis the compound and investigate the thermoelectric properties of this potential candidate.